\documentclass{jpp}

\usepackage{graphicx}
\usepackage{natbib}
\usepackage{epsfig}

\ifCUPmtlplainloaded \else
  \checkfont{eurm10}
  \iffontfound
    \IfFileExists{upmath.sty}
      {\typeout{^^JFound AMS Euler Roman fonts on the system,
                   using the 'upmath' package.^^J}%
       \usepackage{upmath}}
      {\typeout{^^JFound AMS Euler Roman fonts on the system, but you
                   dont seem to have the}%
       \typeout{'upmath' package installed. JPP.cls can take advantage
                 of these fonts, if you use 'upmath' package.^^J}%
      }
  \else
  \fi
\fi

\ifCUPmtlplainloaded \else
  \checkfont{msam10}
  \iffontfound
    \IfFileExists{amssymb.sty}
      {\typeout{^^JFound AMS Symbol fonts on the system, using the
                'amssymb' package.^^J}%
       \usepackage{amssymb}%

      }{}
  \fi
\fi

\ifCUPmtlplainloaded \else
  \IfFileExists{amsbsy.sty}
    {\typeout{^^JFound the 'amsbsy' package on the system, using it.^^J}%
     \usepackage{amsbsy}}
    {}
\fi

\newsavebox{\astrutbox}
\sbox{\astrutbox}{\rule[-5pt]{0pt}{20pt}}

\title[Fluid description of the cooperative scattering..]
{Fluid description of the cooperative scattering of light by
spherical atomic clouds}

\author[N. Piovella, R. Bachelard, Ph.W. Courteille]%
{N.\ns P\ls I\ls O\ls V\ls E\ls L\ls L\ls A$^1$%
  \thanks{Email address for correspondence: nicola.piovella@unimi.it},\ns
R.\ns B\ls A\ls C\ls H\ls E\ls L\ls A\ls R\ls D$^2$\break \and
PH.\ns W.\ns C\ls O\ls U\ls R\ls T\ls E\ls I\ls L\ls L\ls
E\ls$^2$}
\affiliation{$^1$Dipartimento di Fisica, Universit\`a degli Studi
di Milano, Via Celoria 16, Milano I-20133, Italy\\[\affilskip]
$^2$Instituto de F\'{i}sica de S\~{a}o Carlos, Universidade de
S\~{a}o Paulo, 13560-970 S\~{a}o Carlos, SP, Brazil}

\pubyear{2010}
\volume{650}
\pagerange{119--126}
\date{?; revised ?; accepted ?. - To be entered by editorial office}
\begin{document}

\maketitle

\begin{abstract}
When a cold atomic gas is illuminated by a quasi-resonant laser
beam, light-induced dipole-dipole correlations make the scattering
of light a cooperative process. Once a fluid description is
adopted for the atoms, many scattering properties are captured by
the definition of a complex refractive index. The solution of the
scattering problem is here presented for spherical atomic clouds
of arbitrary density profiles, such as parabolic densities
characteristic of ultra-cold clouds. A new solution for clouds
with infinite boundaries is derived, that is particularly useful
for the Gaussian densities of thermal atomic clouds. The presence
of Mie resonances, a signature of the cloud acting as a cavity for
the light, is discussed. These resonances leave their fingerprint
in various observables such as the scattered intensity or in the
radiation pressure force, and can be observed by tuning the
frequency of the incident laser field or the atom number.
\end{abstract}


\section{Introduction}
Ultracold and quantum plasmas have shown an increasing interest
for their analogies with cold atomic systems
\citep{Tito2013,Tito2008}. In the dilute regime, direct particle
interactions can often be neglected and the particles are
interacting only via their common radiation field. Furthermore,
cold atomic systems are often at the borderline between classical
and quantum realms, allowing to investigate in a controlled way
how cooperativity changes when a transition to a quantum regime
occurs. A simple example is provided by a cold atomic cloud
released from a magneto-optical trap and illuminated by a laser
beam \citep{Bienaime10,Bender10}. Such a system manifests many
interesting effects when the atomic cloud loses its granularity
and can be described as a continuous fluid~\citep{Prasad10}. As
size, shape and density of the scatterer vary, a variety of
radiating patterns emerge, well known since the pioneering studies
of Mie on extended dielectric particles \citep{Mie1908,Hulst}.

Although Mie scattering from an uniform dielectric sphere already
represents a rather complex problem, light scattering by cold
atoms adds new features that deserve dedicated
studies~\citep{Bachelard11,Bachelard:EPL12}. Firstly, changing the
laser wavelength allows to tune the light-matter interaction: this
allowed to observe experimentally the transition from single-atom
scattering to Mie scattering by a macroscopic object, when the
collective effects appear~\citep{Courteille10,Bender10}. Secondly,
the weakness of direct inter-particle interactions allow to probe
the forces that light exerts on each individual atom: in
particular, for ultra-cold atomic clouds the pattern of atomic
recoil was shown to contain the history of the interaction, and to
exhibit signatures of anisotropic of the Mie
scattering~\citep{Bachelard:PRA12}. Finally, various geometries,
atomic densities and ordered or disordered structures can be
generated by designing appropriate magnetic and optical traps for
the atoms. This illustrates the flexibility of cold atom
experiments to model and investigate phenomena from various
fields, such as condensed matter and possibly plasmas.

In this paper we present the solution of the Mie scalar scattering
problem for spherical inhomogeneous atomic clouds, with finite and
infinite boundary conditions. Beyond the classical Mie solution
for hard and homogeneous dielectric spheres, it provides a
realistic description of light scattering by dilute atomic clouds.
We discuss the possibility for the cloud to act as a resonant
cavity for the light, and the signatures of this behavior.

\section{The scattering equation}\label{sec:scattering equation}

Let us consider a sample of $N$ cold two-level atoms with resonant
frequency $\omega_a$, linewidth $\Gamma=
d^2\omega_a^3/2\pi\hbar\epsilon_0c^3$ and electric dipole
transition matrix element $d$. When illuminated by an uniform
laser beam with electric field amplitude $E_0$, frequency
$\omega_0$, and wave vector $\mathbf{k}_0 = (\omega_0/c)\hat
\mathbf{e}_z$, their response is described for weak incident
fields by the following coupled equations
\citep{Scully06,Svi08,Svi10,Courteille10}:
\begin{equation}\label{Eqbetaj}
    \frac{\partial{\beta}_j}{\partial t}=
    i\Delta_0\beta_j(t) - i\frac{dE_0}{2\hbar}e^{i\mathbf{k}_0\cdot \mathbf{r}_j}
    -\frac{\Gamma}{2}\sum_{m=1}^N
    \frac{\exp(ik_0|\mathbf{r}_j-\mathbf{r}_m|)}{ik_0|\mathbf{r}_j-\mathbf{r}_m|}\beta_m(t)
\end{equation}
with $\Delta_0=\omega_0-\omega_a$,  $\mathbf{r}_j$ the position of
atom $j$, and $\beta_j$ its complex probability amplitude to be
excited at time $t$. The scattering kernel
\begin{equation}\label{G}
    G(\mathbf{r},\mathbf{r}')=\frac{\exp(ik_0|\mathbf{r}-\mathbf{r}'|)}{ik_0|\mathbf{r}-\mathbf{r}'|}.
    \end{equation}
has a real component that describes the {\it collective} (superradiant) atomic decay \citep{Svi08}, and an
imaginary component that contains the collective Lamb shift due to
light-induced short-range interaction between the atoms~\citep{Friedberg73,Scully09,Scully10}. The
latter becomes significant when the number of atoms in a cubic
optical wavelength, $\rho\lambda^3$, is larger than unity.
 Eq.~(\ref{Eqbetaj}) has been obtained assuming at most
one atom is excited~\citep{Scully06,Courteille10}. However, this
model has also been shown to describe the dynamics of coupled
classical linear oscillators~\citep{Svi10}, so that the many-body
features of cooperative scattering by two-level atoms may be
understood as a classical effect. Remark that this analogy holds
only in the weak excitation limit of the atomic ensemble.
Furthermore, in this approach short-range dipole interactions and
polarization effects are neglected \citep{Friedberg73}.

Neglecting granularity and isolating the self-decaying term
$-(\Gamma/2)\beta_j$, Eq.~(\ref{Eqbetaj}) takes the form of an
integro-differential fluid equation for the complex field
$\beta(\mathbf{r},t)$~\citep{Bachelard:EPL12}:
\begin{equation}\label{Eqbetar}
    \frac{\partial\beta(\mathbf{r},t)}{\partial t}=
    \left(i\Delta_0-\frac{\Gamma}{2}\right)\beta(\mathbf{r},t) - i\frac{dE_0}{2\hbar}e^{i\mathbf{k}_0\cdot \mathbf{r}}
    -\frac{\Gamma}{2}\int d\mathbf{r}'\rho(\mathbf{r}')
    G(\mathbf{r},\mathbf{r}')\beta(\mathbf{r}',t)
\end{equation}
where $\rho(\mathbf{r})$ is the atomic density. At steady-state, Eq.~(\ref{Eqbetar}) yields
\begin{equation}\label{Eqss}
    \int d\mathbf{r}'\rho(\mathbf{r}')
    \frac{\exp(ik_0|\mathbf{r}-\mathbf{r}'|)}{k_0|\mathbf{r}-\mathbf{r}'|}\tilde\beta(\mathbf{r}')=
    -\left(2\delta+i\right)\tilde\beta(\mathbf{r}) + e^{i\mathbf{k}_0\cdot \mathbf{r}}
\end{equation}
where we have introduced the normalized detuning $\delta=\Delta_0/\Gamma$ and excitation field
$\tilde\beta(\mathbf{r})=(\hbar\Gamma/dE_0)\beta(\mathbf{r})$.
 Let us remark that
the kernel of the integral of Eq.~(\ref{Eqss}) is the Green
function for the Helmholtz equation:
\begin{equation}\label{Green}
    \left(\nabla^2+k_0^2\right)\frac{\exp(ik_0|\mathbf{r}-\mathbf{r}'|)}{|\mathbf{r}-\mathbf{r}'|}=
    -4\pi\delta(\mathbf{r}-\mathbf{r}')
\end{equation}
and that $\left(\nabla^2+k_0^2\right)\exp(i\mathbf{k}_0\cdot
\mathbf{r})=0$. Then, applying $(\nabla^2+k_0^2)$ to Eq.~(\ref{Eqss})
we obtain that $\tilde\beta(\mathbf{r})$ satisfies the Helmholtz
equation
\begin{equation}\label{Helm}
    \left[\nabla^2+k_0^2m_0^2(\mathbf{r})\right]\tilde\beta(\mathbf{r})=0
\end{equation}
where $m_0(\mathbf{r})$ is the local refraction index of the atomic
cloud:
\begin{equation}\label{m0}
    m_0^2(\mathbf{r})=1-\frac{4\pi\rho(\mathbf{r})}{k_0^3(2\delta+i)}.
\end{equation}
Hence, the field $\tilde\beta(\mathbf{r})$ propagates as a wave in
the cloud of cold atoms, that acts as a ``classical'' dielectric
medium of index $m_0(\mathbf{r})$. The imaginary part of $m_0$
originates in the single-atom decay term and is responsible for
the diffusive nature of the cloud: it vanishes only in the limit
of far-detuned incident laser.

\section{Scattered intensity and radiation pressure force}

From the local response $\beta(\mathbf{r},t)$ to the
external field, many measurable quantities can be derived, such as the scattered intensity and the radiation
pressure force~\citep{Courteille10}. The scattered field is~\citep{Bachelard11}:
\begin{equation}\label{Es}
    E_s(\mathbf{r},t)=-E_0
    \int d\mathbf{r}'\rho(\mathbf{r}')
    \frac{\exp(ik_0|\mathbf{r}-\mathbf{r}'|)}{k_0|\mathbf{r}-\mathbf{r}'|}\tilde\beta(\mathbf{r}',t)
\end{equation}
and the far-field scattered intensity, at distance $r$ and direction
$(\theta,\phi)$ with respect to the $z$-axis, reads
\begin{equation}\label{Is}
    I_s(r,\theta,\phi)=c\epsilon_0\frac{E_0^2}{k_0^2r^2
    }\left[
    N\langle|\tilde\beta(\mathbf{r})|^2\rangle
    +N^2|s(\mathbf{k})|^2\right]
\end{equation}
where
\begin{equation}\label{aveb2}
    \langle|\tilde\beta(\mathbf{r})|^2\rangle=\frac{1}{N}\int d\mathbf{r}'\rho(\mathbf{r}')
    |\tilde\beta(\mathbf{r}')|^2
\end{equation}
and $s(\mathbf{k})$ is the structure factor,
\begin{equation}\label{sk}
    s(\mathbf{k})=\frac{1}{N}\int d\mathbf{r}\rho(\mathbf{r})\tilde\beta(\mathbf{r}) e^{-i\mathbf{k}\cdot
    \mathbf{r}}
\end{equation}
with
$\mathbf{k}=k_0(\sin\theta\cos\phi,\sin\theta\sin\phi,\cos\theta)$.
The first term of Eq.(\ref{Is}) corresponds to the incoherent and isotropic contribution, and is proportional to $N$, whereas the second term
is the superradiant, strongly directional, scattered
intensity, and it is proportional to $N^2$. Hence, the cooperation of an increasing number of atoms to scatter light
leads to a more coherent and focused emission -- superradiance dominates over single-atom scattering.

Another observables of interest is the radiation pressure force,
that is the net force that the light exerts on the atoms. It is
the sum of two contributions, the first due to absorption of
photons from the incident field and the second one due to their
spontaneous re-emission along any direction
$(\theta,\phi)$~\citep{Courteille10,Bachelard11}. The force on the
cloud center-of-mass and along the $z$-axis reads
\begin{equation}\label{F}
    \langle F_{z}\rangle=\langle F_{z}^a\rangle+\langle F_{z}^e\rangle,
\end{equation}
where
\begin{eqnarray}
    \langle F_{z}^a\rangle&=&-2\pi\epsilon_0\frac{E_0^2}{k_0^2}\textrm{Im}(s(\mathbf{k}_0)),\label{Fa}\\
     \langle F_{z}^e\rangle&=&
         -\epsilon_0\frac{E_0^2N}{2k_0^2}\int d\Omega_k \cos\theta
    |s(\mathbf{k})|^2,\label{Fe}
\end{eqnarray}
with $d\Omega_k=d\phi\,d\theta\sin\theta$ the elementary solid
angle. The absorption force $\mathbf{F}^a$ is directed along the
$z$-axis and it is proportional to the incident intensity,
$I_0=\epsilon_0 cE_0^2$ since it corresponds to the absorption of
laser photons. The emission force $\mathbf{F}^e$ is directed along
$\mathbf{k}$, and its $z$-component is obtained by averaging over
the solid angle the product between the emission probability of
the photons $|s(\mathbf{k})|^2$ and the projection of their
momentum on the $z$-axis, hence the factor $\cos\theta$. The
integration over the solid angle $\Omega_k$ in Eq.~(\ref{Fe}) is
performed using the identity
\begin{equation}\label{j1}
\int d\Omega_k\cos\theta
e^{i\mathbf{k}\cdot(\mathbf{r}-\mathbf{r}')}=4\pi
i\frac{z-z'}{|\mathbf{r}-\mathbf{r}'|}j_1(k_0|\mathbf{r}-\mathbf{r}'|)=
-\frac{4\pi i}{k_0}\frac{\partial}{\partial
z}j_0(k_0|\mathbf{r}-\mathbf{r}'|),
\end{equation}
where $j_0(x)=\sin(x)/x$ and $j_1(x)=\sin(x)/x^2-\cos(x)/x$ are
the zero-th and first-order spherical Bessel functions. Using Eq.~(\ref{j1}), the
emission force is written as
\begin{eqnarray}
     \langle F_{z}^e\rangle&=&
         2\pi\epsilon_0\frac{E_0^2}{k_0^2N}\textrm{Re}\left[ \int d\mathbf{r}\rho(\mathbf{r})
         \tilde\beta^*(\mathbf{r})\frac{\partial}{\partial z}\int d\mathbf{r}'\rho(\mathbf{r}')
          \frac{\exp(ik_0|\mathbf{r}-\mathbf{r}'|)}{k_0|\mathbf{r}-\mathbf{r}'|}\tilde\beta(\mathbf{r}')
         \right]\label{Fe2}
\end{eqnarray}
Combining Eqs.~(\ref{Fa}) and (\ref{Fe2}) allows to write the total force as
\begin{equation}\label{Ffield}
   \langle F_z\rangle =\frac{1}{N}\int d\mathbf{r}\rho(\mathbf{r})
    \textrm{Re}\left\{-d\beta^*(\mathbf{r})\nabla_z\left[E_0 e^{i\mathbf{k}_0\cdot \mathbf{r}}
    -E_0\int d\mathbf{r}'\rho(\mathbf{r}') \frac{\exp(ik_0|\mathbf{r}-\mathbf{r}'|)}
    {k_0|\mathbf{r}-\mathbf{r}'|}\tilde\beta(\mathbf{r}')\right]\right\}.
\end{equation}
Since the term in the square bracket is the total electric field
$E_{t}(\mathbf{r})=E_0 e^{i\mathbf{k}_0\cdot
\mathbf{r}}+E_s(\mathbf{r})$, i.e., the sum of the incident and
scattered field given by Eq.(\ref{Es}), the force on the center of
mass appears as the average over the local force on the atoms:
\begin{equation}\label{F1atom}
    \mathbf{F}=-d\textrm{Re}(\beta^*\nabla_\mathbf{r}E_t).
\end{equation}
Hence, we recover the well-known expression of the force that a
light field exerts on an atom~\citep{Gordon80}, though a crucial
difference is that $E_t$ here contains the self-radiated field of
the cloud. Note also that in Eq.~(\ref{F1atom}), we have extended
our expression (\ref{Ffield}) to every spatial direction.

\section{Mie scattering}\label{sec:stationary solution}

\begin{figure}
\center
\begin{tabular}{cc}
  \includegraphics[width=6cm]{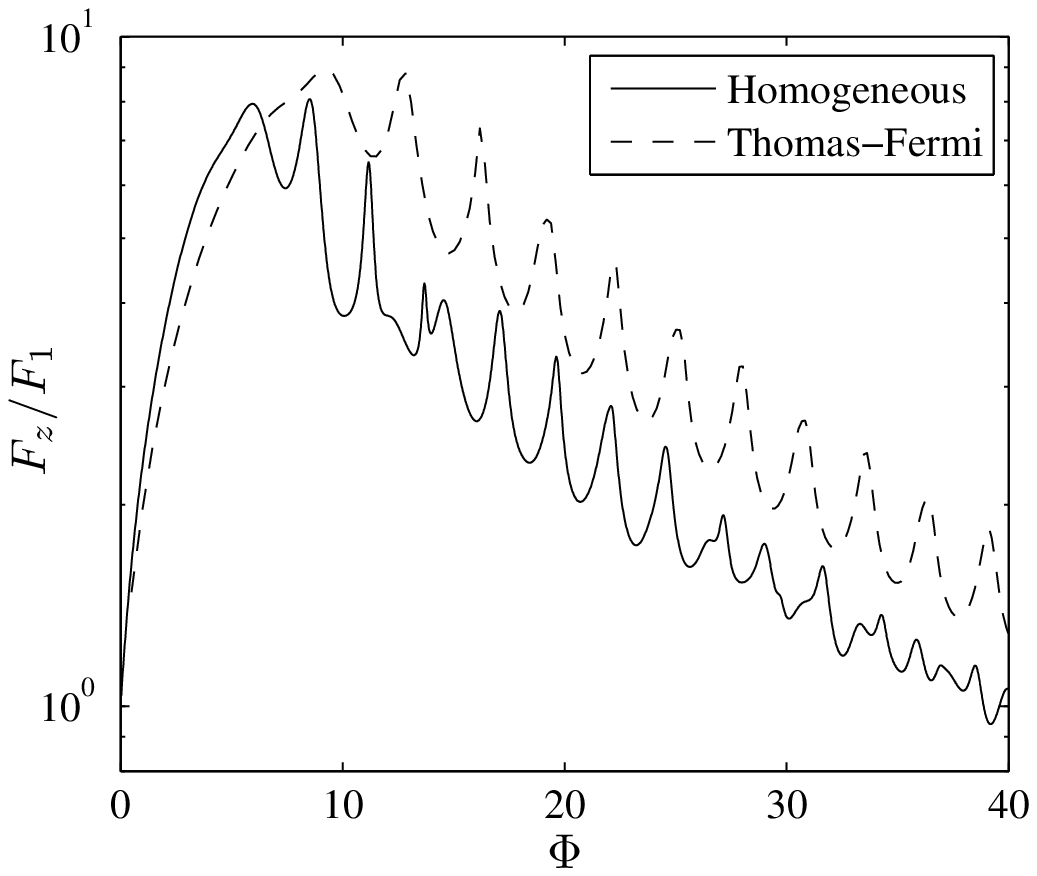}&\includegraphics[width=6cm]{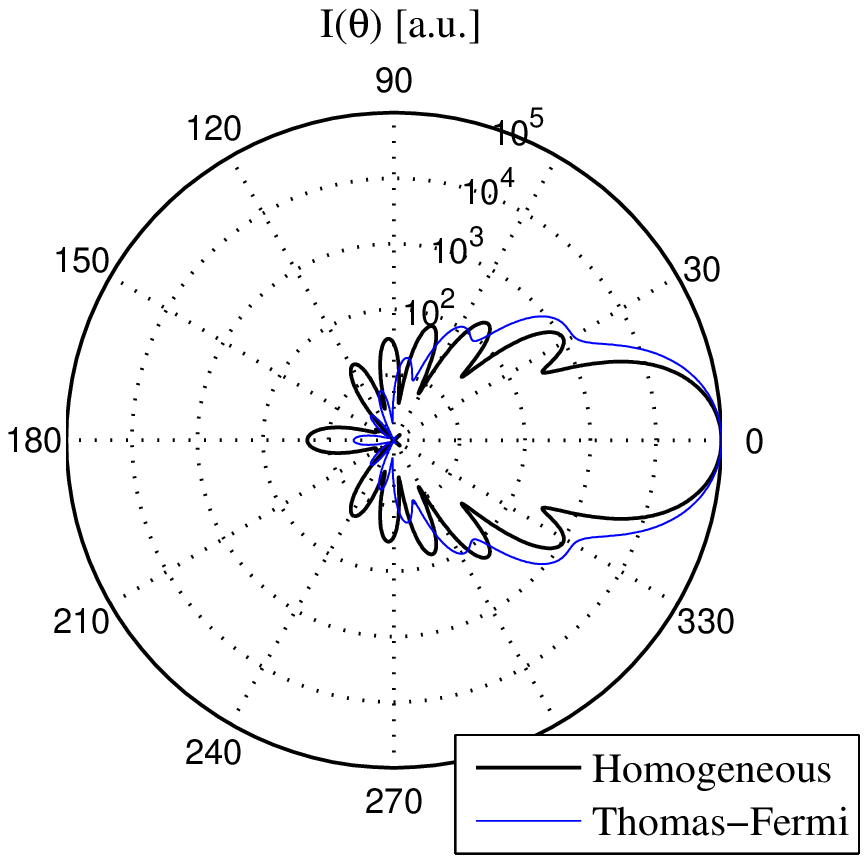}
  \end{tabular}
  \caption{Left: Radiation pressure force as a function of the phase-shift $\Phi$ for homogeneous (plain line) and
  Thomas-Fermi (dashed line) distribution. Simulations realized for a cloud of size $k_0R=10$, a detuning $\delta=-50$,
  and varying the number of atoms $N$. The force is normalized by the single atom force $F_1=2\pi\epsilon_0(E_0/k_0)^2/(1+4\delta^2)$,
  that is observed in absence of collective effects. Right: Radiation pattern $I(\theta)$ for homogeneous (thick black line)
  and Thomas-Fermi (blue thin line) distributions. Simulations realized for a cloud for size $k_0R=10$, detuning $\delta=-50$ and $N=6100$ atoms.}
\label{fig1}
\end{figure}

As discussed before, the knowledge of the excitation probability
$\beta(\mathbf{r})$ is the key to understanding the radiation
pattern and the forces exerted on the atomic cloud. Despite the
three-dimensional nature of the problem, an analytical solution
exists for simple geometries. We here focus on clouds with
spherical symmetry $\rho(r)$, for which the eigenmodes of the wave
equation have the form $u_n(r)P_{n}(\cos\theta)$, with $P_n(x)$
the $n$-th Legendre polynomial and
 $u_n(r)$ a radial mode that satisfies~\citep{Bachelard:EPL12}:
\begin{equation}\label{un}
    u_n''(r)+2\frac{u_n'(r)}{r}+\left[k_0^2m_0^2(r)-\frac{n(n+1)}{r^2}\right]u_n(r)=0.
\end{equation}
Because of the rotational symmetry, no dependence on $\phi$ is present. The excitation field is then decomposed as a sum of partial waves:
\begin{equation}\label{betan}
    \tilde\beta(r,\theta)=\sum_{n=0}^\infty (2n+1)\beta_nu_n(r)P_n(\cos\theta),
\end{equation}
whereas the incident wave is
decomposed as:
\begin{equation}\label{inc}
    e^{i\mathbf{k}_0\cdot \mathbf{r}}=\sum_{n=0}^\infty
    i^n(2n+1)j_n(k_0r)P_n(\cos\theta).
\end{equation}
Finally, the scattering kernel is expanded in
partial waves, using the identity
\begin{eqnarray}\label{kernel}
    \frac{\exp(ik_0|\mathbf{r}-\mathbf{r}'|)}{k_0|\mathbf{r}-\mathbf{r}'|}&=&4\pi i
    \sum_{n=0}^\infty\sum_{m=-n}^nY_{n,m}(\hat r)Y_{n,m}^*(\hat r')
    \left\{
    \begin{array}{c}
     j_n(k_0r')h^{(1)}_n(k_0r)\quad \textrm{for}\quad r>r'\\
     j_n(k_0r)h^{(1)}_n(k_0r')\quad \textrm{for}\quad r<r' \\
    \end{array}%
    \right.
\end{eqnarray}
where $\hat r$ and $\hat r'$ are unit vectors in the directions of
$\mathbf{r}$ and $\mathbf{r}'$, $Y_{n,m}$ the spherical harmonics
(in particular,
$Y_{n,0}(\theta,\phi)=\sqrt{(2n+1)/4\pi}P_n(\cos\theta)$), and
 $j_{n}(z)$ and $h^{(1)}_{n} (z)$ are the spherical Bessel functions.
 Inserting Eqs.~(\ref{betan}),~(\ref{inc}) and~(\ref{kernel}) into Eq.~(\ref{Eqss})
 and projecting on the orthogonal basis of the Legendre polynomial, we obtain the relation
\begin{equation}\label{eqfn}
    j_n(k_0r)=(2\delta+i)\beta_n u_n(r)+\beta_n f_n(r),
\end{equation}
where
\begin{equation}\label{fn}
    f_n(r)=4\pi i \int_0^\infty dr' r'^2\rho(r')u_n(r')\left\{
    \begin{array}{c}
     j_n(k_0r')h^{(1)}_n(k_0r)\quad \textrm{for}\quad r>r'\\
     j_n(k_0r)h^{(1)}_n(k_0r')\quad \textrm{for}\quad r<r' \\
    \end{array}%
    \right.
\end{equation}
Once specified the form for $\rho(r)$, the solution for the radial
mode $u_n$ can be derived, and the amplitude of each partial wave
$\beta_n$ calculated from Eqs.~(\ref{eqfn}) and (\ref{fn}).
Assuming a finite cloud of radius $R$, such that $\rho(r)=0$ for
$r>R$, it reads~\citep{Bachelard:EPL12}:
\begin{equation}\label{bnR}
    \beta_n=\frac{j_n(k_0R)}{(2\delta+i)u_n(R)+i\lambda_n h^{(1)}(k_0R)}
\end{equation}
with
\begin{equation}\label{lambdagen}
    \lambda_n=4\pi\int_0^\infty dr r^2\rho(r)j_n(k_0r)u_n(r),
\end{equation}
and where we have used that
\begin{equation}\label{fnR}
    f_n(R)=4\pi i h^{(1)}(k_0R)\int_0^R dr' r'^2\rho(r')u_n(r')j_n(k_0r')=i\lambda_n h^{(1)}(k_0R).
\end{equation}
The cloud structure factor is then easily derived~\citep{Bachelard11}:
\begin{equation}\label{sk2}
    s(\mathbf{k})=\frac{1}{N}\sum_{n=0}^\infty
    (2n+1)\lambda_n\beta_nP_n(\cos\theta),
\end{equation}
as well as the isotropic radiation contribution:
\begin{equation}\label{aveb2bis}
    \langle|\tilde\beta(\mathbf{r})|^2\rangle= \frac{1}{N}\sum_{n=0}^\infty
    (2n+1)\breve{\lambda}_n|\beta_n|^2,
\end{equation}
where
\begin{equation}\label{lambdabreve}
    \breve{\lambda}_n=4\pi\int_0^\infty dr r^2\rho(r)|u_n(r)|^2.
\end{equation}
The problem of scattering by a homogeneous spherical dielectrics
was initially investigated by Gustav Mie~\citep{Mie1908}, and it
was later generalized to homogeneous ellipsoids. In the case of a
homogeneous spherical atomic cloud of $N$ atoms and radius $R$,
the index is $m_0 =\sqrt{1-3N/(k_0R)^3(2\delta +i)}$, and the
radial solutions of the Helmholtz equation is
$u_n(r)=j_n(m_0k_0r)$. Using the special properties of the Bessel
functions, $\lambda_n$ is explicitly calculated:
\begin{equation}\label{lambdanhomo}
    \lambda_n=(2\delta +i)(k_0R)^2\left[m_0j_{n-1}(m_0k_0R)j_n(k_0R)-j_{n-1}(k_0R)j_n(m_0k_0R)\right],
\end{equation}
and the complex amplitude of the $n$-th partial wave is deduced:
\begin{equation}\label{bnhomo}
    \beta_n=\frac{j_n(k_0R)}{(2\delta+i)j_n(m_0k_0R)+i\lambda_n h^{(1)}(k_0R)}.
\end{equation}
As can be observed in Fig.~\ref{fig1}, when the optical density is
tuned -- here varying the number of particles at fixed volume and
detuning, the force that pushes the cloud exhibits some
oscillations. These resonances correspond to the cloud acting as a
resonant cavity for the light, and are best understood by
introducing the phase-shift experienced by the light inside the
cloud: $\Phi=\int (m_0(0,0,z)-1) dz$. When this phase-shift is a
multiple of $\pi$, the cavity that the cloud forms is at resonance
with the light, so a greater amount of light is stored in the
atomic cloud and the radiation pressure force
increases~\citep{Bachelard:EPL12}.

Yet the spiky structure of the force for homogeneous clouds hides
more complex resonances. It is known that the sharp index
interface between a dielectric and, say, vacuum allows for surface
modes to propagate. These are called whispering gallery
modes~\citep{Nussenzveig92}. They are characterized by sharp
resonances whose number, positions and amplitudes strongly depend
on the cloud characteristics, making them a powerful tool to
characterize the scatterers~\citep{Oraevsky02}.

This exact solution of the scattering problem as an infinite
series of partial waves is reminiscent of Mie
solution~\citep{Mie1908,Hulst}, although our approach is slightly
different: Mie used continuity equations for the tangential field
components at the dielectric boundary, while we adopted an
integral formulation of the problem. The two solutions are
nevertheless formally the same~\citep{Bachelard:EPL12}, as far as
finite clouds are concerned. Let us also remark that our solution
is more general, since it holds for any radial solution $u_n(r)$
of the Helmholtz equation, i.e., it applies to any spherical cloud
with arbitrary density. It is thus of particular interest for
atomic clouds where the traps in general generate inhomogeneous
density profiles.

\section{Ultra-cold clouds}

Ultracold clouds of fermionic species have been successfully described by quadratic profiles of density,
following the pioneering work of Llewellyn Thomas and Enrico Fermi on distributions of electrons~\citep{Thomas27,Fermi27}.
These clouds exhibit a parabolic density $\rho(r)=(5N/2V)[1-r^2/R^2]$, with $R$ the radius of the cloud and $V =4\pi R^3/3$
its volume, and consequently a spatially-dependent refraction index:
\begin{equation}
m^2_0(r)=m^2_c +\gamma^2r^2,
\end{equation}
with $m_c = \sqrt{1-(15/2)N/(k_0R)^3(2\delta +i)}$ the index in
the core of the sample, and $\gamma^2 =(15/2)N/(k_0^3R^5)(2\delta
+i)$.  Using the substitution $u_n(r)=r^{-3/2}w_n(x)$, with $x=
\gamma r^2/2$, one can show that $w_n(x)$ satisfies the Coulomb
wave equation, well known in nuclear physics~\citep{Martin02},
whose solutions are the Coulomb wave functions
$w_n(x)=F_{n/2-1/4}(-m_c^2/4\gamma,x)$. We get the following
partial wave expansion for the excitation field:
\begin{equation}\label{unCoul}
\tilde\beta(r,\theta)=\sum_{n=0}^\infty (2n+1)
\frac{\beta_n}{r^{3/2}}F_{\frac{n}{2}-\frac{1}{4}}\left(-\frac{m_c^2}{4\gamma},\frac{\gamma
   r^2}{2}\right)P_n(\cos\theta).
\end{equation}
Apart from the fact that studying the scattering by atomic clouds
with a quadratic density rather than an homogeneous one is a much
more realistic approximation, the former profile also yields
different physics. Indeed, as can be observed in Fig.~\ref{fig1},
the spiky structure of the whispering gallery modes disappears,
and only the regular oscillations of the longitudinal cavity
modes, such that the longitudinal phase-shift in the cloud is a
multiple of $\pi$, survive. Indeed, because the change in
refractive index is much smoother in clouds with a Thomas-Fermi
distribution than with a homogeneous one, the surface modes cannot
propagate anymore.

\begin{figure}
\center
\begin{tabular}{cc}
  \includegraphics[width=6cm]{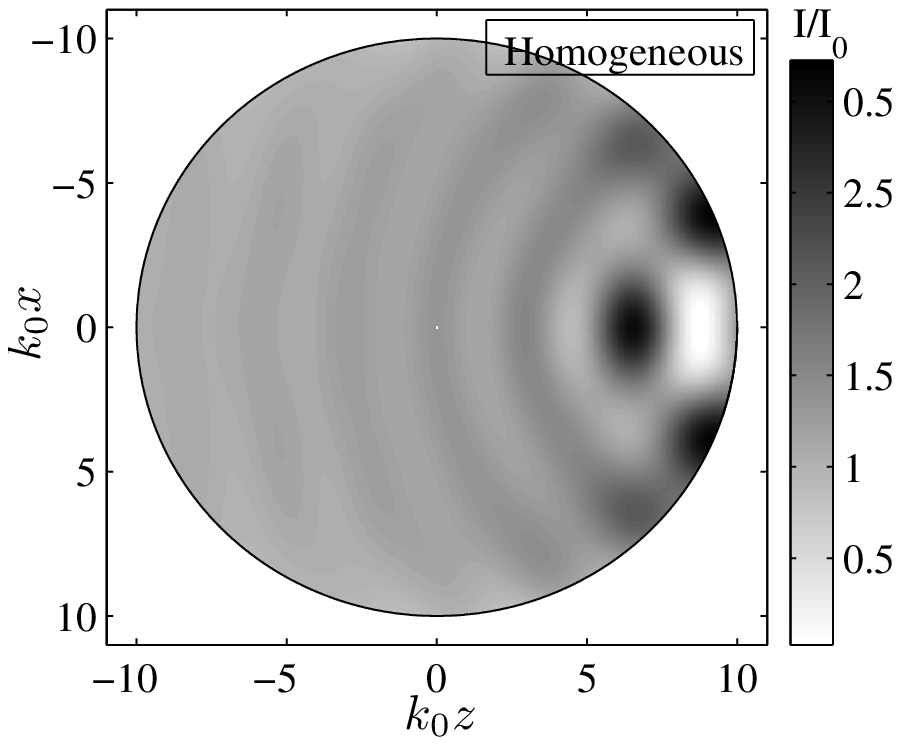}&\includegraphics[width=6cm]{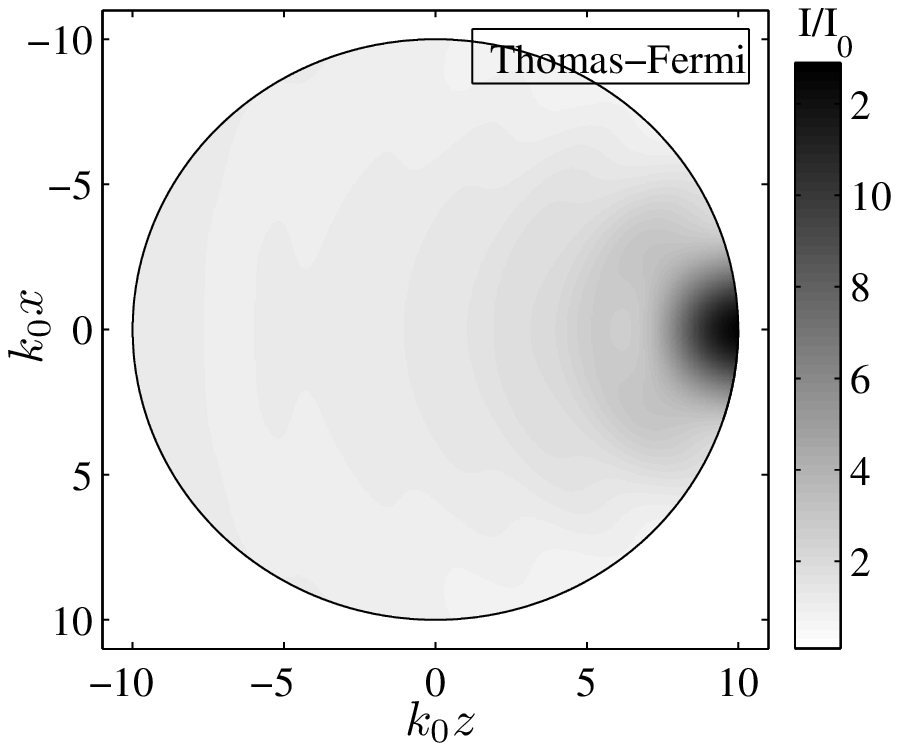}
  \end{tabular}
  \caption{Light intensity in the cloud with homogeneous (left) and Thomas-Fermi (right) distributions, in the plane $y=0$.
  Simulations realized for a cloud of size $k_0R=10$, a detuning $\delta=-50$ and $N=6100$ atoms. The intensity is normalized to the
  laser intensity $I_0$.}
\label{fig2}
\end{figure}

This analysis is confirmed by the profiles of light intensity in
the cloud, see Fig.~\ref{fig2}. One can observe that homogeneous
spheres exhibit off-axis surface modes (left picture) -- this
feature was checked for various set of parameters. This surface
modes create new resonances that compete with the longitudinal
ones, causing the spiky structure observed in Fig.~\ref{fig1} for
the center-of-mass force. These off-axis modes appear much weaker
in clouds with Thomas-Fermi distribution (right picture in
Fig.~\ref{fig2}).

\section{Thermal clouds}

Finally, we discuss the Mie scattering solution in clouds with
boundaries extending until $r=\infty$. For instance, thermal
clouds with a Maxwell-Boltzmann velocity distribution have a
Gaussian density profile, so the Mie scattering solution that
makes use of finite boundary conditions cannot be used
straightforwardly. Here we propose an alternative derivation that
holds as well for such distributions with boundaries at infinite.
To our knowledge, it is the first solution to the Mie scattering
problem with infinite boundaries.

Starting from Eq.~(\ref{eqfn}), the amplitude $\beta_n$ of the $n$-th partial wave can be obtained by
multiplying both terms by $4\pi r^2\rho(r)j_n(k_0r)$ and integrating over $r$:
\begin{equation}\label{bnrr}
    \beta_n=\frac{A_n}{(2\delta+i)\lambda_n+B_n},
\end{equation}
where we have introduced:
\begin{eqnarray}
  A_n &=&  4\pi\int_0^\infty dr r^2\rho(r) j_n^2(k_0 r),\label{An}\\
  B_n &=&  4\pi\int_0^\infty dr r^2\rho(r) j_n(k_0 r)f_n(r).\label{Bn}
\end{eqnarray}
Using Eq.(\ref{fn}), the coefficient $B_n$ can be written as
\begin{eqnarray}
  B_n =  16\pi^2 i\int_0^\infty dr r^2\rho(r) j_n(k_0 r)
  &\times&\left\{h_n^{(1)}(k_0r)\int_0^r dr'
  r'^2\rho(r')u_n(r')j_n(k_0r')\right.\nonumber\\
  &+&\left.
  j_n(k_0r)\int_r^\infty dr'
  r'^2\rho(r')u_n(r')h_n^{(1)}(k_0r')\right\}.
  \label{Cnb}
\end{eqnarray}
Using the identities (\ref{A5}) and (\ref{A6}) of Appendix~\ref{appA}, and the identity
$j_n(z){h_n^{(1)}}'(z)-h_n^{(1)}(z)j_n'(z)=i/z^2$, one can show that
\begin{equation}\label{Cn2}
    B_n=(2\delta+i)(-\lambda_n+iA_n D_n)
\end{equation}
with
\begin{equation}\label{Dn}
    D_n=\lim_{r\rightarrow\infty}\left\{
    r^2\left[
    h_n^{(1)}(k_0r)u_n(r)-u_n(r){h_n^{(1)}}'(k_0r)
    \right]
    \right\}.
\end{equation}
One deduces that:
\begin{equation}\label{betaInf}
    \beta_n=\frac{1}{i(2\delta+i)D_n}=\frac{1}{2\delta+i}\lim_{r\rightarrow\infty}
    \left\{
    \frac{1}{ir^2\left[h_n^{(1)}(k_0r)u_n'(r)-u_n(r){h_n^{(1)}}'(k_0r)\right]}
    \right\}.
\end{equation}
As for the eigenvalues $\lambda_n$ of the scattering problem, they
read:
\begin{equation}\label{lambdaInf}
    \lambda_n=4\pi\int_0^\infty dr
    r^2\rho(r)j_n(k_0r)u_n(r)=(2\delta+i)\lim_{r\rightarrow\infty}
    \left\{
    r^2[j_n(k_0r)u_n'(r)-u_n(r)j_n'(k_0r)]
    \right\}.
\end{equation}
Thus, the scattering coefficients are determined by the value of
the radial modes $u_n(r)$ and of their first derivative at
$r\rightarrow\infty$.

\section{Discussion}

We have here discussed the light scattering by a macroscopic
atomic cloud, when the atoms cooperate to scatter the light
superradiantly. The cloud was considered as a fluid, i.e., the
point-like nature of the microscopic scatterers was neglected, and
an analytical solution was then derived for spherical geometries,
where the excitation field inside the cloud is developed as the
sum of partial waves. Although we generalized it to arbitrary
spherical densities, this technique is formally equivalent to Mie
scattering, where continuity equations are used at the boundaries
of the scattering medium. Furthermore, our technique allowed for
the derivation of a solution to the Mie problem for clouds with
infinite boundaries. An accurate treatment of the decay of the
density profile is crucial to understand if some special
resonances, such as whispering gallery modes, may exist or not. It
is also important in the context of atomic clouds or plasmas
where, differently from solid dielectrics, the densities are
usually strongly non-homogeneous.

\appendix

\section{}\label{appA}
The function $u_n(r)$ is a solution of the differential equation
(\ref{un}):
\begin{equation}\label{A1}
    u_n''+2\frac{u_n'}{r}+\left[k_0^2m_0^2(r)-\frac{n(n+1)}{r^2}\right]u_n=0.
\end{equation}
where $m_0^2(r)=1-4\pi\rho(r)/k_0^3(2\delta+i)$. Defining
$v_n(r)=ru_n(r)$, Eq.~(\ref{A1}) becomes
\begin{equation}\label{A2}
    v_n''+\left[k_0^2m_0^2(r)-\frac{n(n+1)}{r^2}\right]v_n=0.
\end{equation}
For $m_0(r)=1$, the solution is $q_n(r)=rj_n(k_0r)$. Introducing
$P(r)=k_0^2m_0^2(r)-n(n+1)/r^2$ and $Q(r)=k_0^2-n(n+1)/r^2$, we
get $v_n''+Pv_n=0$ and $q_n''+Qq_n=0$, so that
\begin{equation}\label{A3}
    \int dr (Q-P)v_nq_n=q_n v_n'-v_n q_n'.
\end{equation}
Since
\begin{equation}\label{A4}
    (Q-P)v_nq_n=k_0^2(1-m_0^2)v_nq_n=\frac{4\pi\rho(r)}{k_0(2\delta+i)}r^2
    u_n(r)j_n(k_0r),
\end{equation}
we obtain the indefinite integral
\begin{equation}\label{A5}
    \frac{4\pi}{k_0}\int^r dr'
    r^{\prime2}\rho(r')u_n(r')j_n(k_0r')=(2\delta+i)r^2\left\{j_n(k_0r)u_n'(r)-u_n(r)j_n'(k_0r)
    \right\}.
\end{equation}
In a similar way we obtain
\begin{equation}\label{A6}
    \frac{4\pi}{k_0}\int^r dr'
    r^{\prime2}\rho(r')u_n(r')h_n^{(1)}(k_0r')=(2\delta+i)r^2\left\{h_n^{(1)}(k_0r)u_n'(r)-u_n(r){h_n^{(1)}}'(k_0r)
    \right\}.
\end{equation}

\bibliographystyle{jpp}

\bibliography{jpp-instructions-Piovella}

\end{document}